\input psfig.sty
\documentstyle{cupconf}

\ifoldfss
\else
  \ifnfssone
    \newmathalphabet{\mathit}
      \addtoversion{normal}{\mathit}{cmr}{m}{it}
      \addtoversion{bold}{\mathit}{cmr}{bx}{it}
    \newmathalphabet{\mathcal}
      \addtoversion{normal}{\mathcal}{cmsy}{m}{n}
    \else
    \ifnfsstwo
    \fi
  \fi
\fi

\def\hexnumber#1{\ifcase#1 0\or1\or2\or3\or4\or5\or6\or7\or8\or9\or
 A\or B\or C\or D\or E\or F\fi }

\makeatletter
\ifx\CUP@mtlplain@loaded\undefined
\else
\fi
\makeatother
 \makeatletter
 \ifx\CUP@mtlplain@loaded\undefined
   \font\tenbmi=cmmib10 at 10pt
   \font\sevenbmi=cmmib10 at 7pt
   \font\fivebmi=cmmib10 at 5pt

   \newfam\bmifam
   \textfont\bmifam=\tenbmi
   \scriptfont\bmifam=\sevenbmi
   \scriptscriptfont\bmifam=\fivebmi
   
 \fi
 \makeatother
\ifnfsstwo

\fi
\ifnfssone

\fi
\ifoldfss

\fi

\mathchardef\varLambda="0103

\makeatletter
\ifx\CUP@mtlplain@loaded\undefined
\else
\fi
\makeatother
\makeatletter
\ifx\CUP@mtlplain@loaded\undefined
  \font\tenbms=cmbsy10
  \font\sevenbms=cmbsy10 at 7pt
  \font\fivebms=cmbsy10 at 5pt
  \newfam\bmsfam
  \textfont\bmsfam=\tenbms
  \scriptfont\bmsfam=\sevenbms
  \scriptscriptfont\bmsfam=\fivebms

  \edef\bsy@{\hexnumber\bmsfam}
  \mathchardef\bnabla="0\bsy@72
\fi
\makeatother
\title[
AGN and starbursts at high redshift: High resolution
               EVN radio observations of the Hubble Deep Field.
]
{
AGN and starbursts at high redshift: EVN observations of the HDF.
}

\author[ ]{%
M.\ls A.\ns G\ls A\ls R\ls R\ls E\ls T\ls T$^1$, 
T.\ls W.\ls B.\ns M\ls U\ls X\ls L\ls O\ls W, 
S.\ls T.\ns G\ls A\ls R\ls R\ls I\ls N\ls G\ls T\ls O\ls N, 
W.\ns A\ls L\ls E\ls F\ls, A.\ns A\ls L\ls B\ls E\ls R\ls D\ls I\ls,
H.\ls J.\ns van\ls L\ls A\ls N\ls G\ls E\ls V\ls E\ls L\ls D\ls E\ls,  
T.\ns  V\ls E\ls N\ls T\ls U\ls R\ls I\ls, A.\ls G.\ns P\ls O\ls L\ls
A\ls T\ls I\ls D\ls I\ls S\ls, K.\ls I.\ns K\ls E\ls L\ls L\ls E\ls
R\ls M\ls A\ls N\ls N,\ls W.\ls A.\ns B\ls A\ls A\ls N\ls, A.\ns K\ls U\ls S\ls, 
A.\ls M.\ls S.\ns R\ls I\ls C\ls H\ls A\ls R\ls D\ls S,\ls
               P.\ls N.\ns W\ls I\ls L\ls K\ls I\ls N\ls S\ls O\ls N\ls}

\affiliation{$^1$Joint Institute for VLBI in Europe, Postbus 2, 7990 AA
  Dwingeloo, The Netherlands}

\begin{document}
\ifnfssone
\else
  \ifnfsstwo
  \else
    \ifoldfss
      \let\mathcal\cal
      \let\mathrm\rm
      \let\mathsf\sf
    \fi
  \fi
\fi

\maketitle

\begin{abstract}
  
  VLBI now has the capability of imaging relatively large
  fields-of-view ($\sim 5$ arcmin$^2$) with sub-mJy detection limits.
  In principle such observations can distinguish between starburst and
  AGN activity in cosmologically distant sources. We present the first
  deep, wide-field European VLBI Network (EVN) 1.6 GHz observations of
  faint radio sources in the Hubble Deep Field (HDF) region. The EVN
  clearly ($> 5\sigma$) detects 2 radio sources in a field that
  encompasses the HDF and part of the Hubble Flanking Fields (HFF).
  The sources detected are: VLA~J123644+621133 (a z=1.01,
  low-luminosity FR-I radio source located within the HDF itself) and
  VLA~J123642+621331 (a dust enshrouded, optically faint, z=4.424
  starburst system). A third radio source, J123646+621404, is detected
  at the $4\sigma$ level. The observations presented here, suggest
  that the radio emission in the dusty starburst, VLA~J123642+621331,
  arises mainly from AGN activity, suggesting that in this system, both
  AGN and starburst activity co-exist. The prospects for future VLBI
  deep fields are briefly discussed.

\end{abstract}

\firstsection 
\section{Introduction and Data Analysis}

With the recent implementation of both wide-field (\cite{G99}) and
phase-referencing techniques ({\it e.g.} \cite{BC95}), VLBI now has the
capability of imaging relatively large fields-of-view ($\sim 5$
arcmin$^2$) with sub-mJy detection limits. It is thus possible for VLBI
to begin to contribute to our understanding of the nature of the faint,
sub-mJy and microJy radio source population.  Hitherto such
investigations have been limited to connected element interferometers
such as the VLA (\cite{R98}), MERLIN (\cite{M99}) and most recently the WSRT
(\cite{G00a}). These observations show that the radio structure of
these faint sources is usually sub-galactic in size, the emission has a
steep spectrum and identifications via optical, infrared and sub-mm
instruments suggest the vast bulk can be identified with cosmologically
distant starburst galaxies (often with distorted optical morphologies).
These deep radio surveys are thus crucially important to some
hot-topics of the day, including galaxy formation and evolution in the
early universe. There are many important questions in this area that
can benefit from high sensitivity VLBI observations. In particular,
there is accumulating evidence that both starburst and AGN activity (in
the early universe) are related to galaxy interactions and mergers. It
seems possible that both starburst and AGN activity often co-exist in
the same system.

VLBI, with its ability to provide milliarcsecond angular resolutions
can in principle distinguish between the two phenomena. In starburst
galaxies the radio emission is believed to arise from relatively
extended star formation processes (such as radio emission from relic
SNR, young SNR, HII regions) whereas in AGN the compact radio emission
is usually associated with a powerful central engine, regulated by a
super-massive black hole. Fortunately, the starburst and AGN phenomena
present (with high angular resolution) very different radio morphology
(c.f. Pedlar and Gabuzda - these proceedings).

In this paper, we present the first VLBI observations of faint radio
sources in the Hubble Deep Field (\cite{W96}). The observations were
made using the European VLBI Network (EVN) utilising the new, high
sensitivity MkIV Data Acquisition System.  EVN Observations of the HDF
region were made on 12-14 November 1999 at 1.6 GHz. A total of 32 hours
observing time was split into two, 16 hour runs. Seven EVN telescopes
successfully participated, including the (DSN/NASA) 70-m telescope at
Robledo, Spain.  The observations were made in phase-reference mode and
correlated at the NRAO VLBA correlator in Socorro, USA. In order to
image the entire HDF and part of the Flanking Fields (a region spanning
$\sim 6$ arcmin$^{2}$), the data were maintained in their original form
(1 second integrations, $512 \times 0.125$~MHz channels). After
applying the standard VLBI calibration techniques, these data were
Fourier transformed to the image plane, and six dirty maps were
simultaneously generated, targeting known areas of radio emission
detected via earlier 1.4~GHz MERLIN observations (with MERLIN peak flux
$> 60\mu$Jy/beam). In order to check the robustness of our data and
imaging routines, one of the six fields was centred on an ``empty''
region of radio sky located within the HDF.  It should be noted that
the brightest source in the HDF has a total WSRT flux density of only
1.6~mJy (\cite{G00a}) -- the EVN observations (with a beam size area that
is 1 million times smaller than the WSRT) might best be considered as
the {\it first VLBI observations of blank radio sky!}

The naturally weighted, {\it dirty} images of the HDF and HFF sources
are shown in Fig~1.  The images are restored with a Gaussian elliptical
beam ($26\times20$ mas in PA$=-31^{\circ}$).  The r.m.s.  noise levels
(as measured by IMEAN over the entire image) range from $41\mu$Jy/beam
(in a field with a clear detection) to $33\mu$Jy/beam (in a field with
no detection).

\section{EVN Detections in the central HDF region} 

At the $5\sigma$ detection level, the EVN observations clearly detect
VLA~J123642+621331 (at the phase centre) and VLA~J123644+621133
(located within the HDF itself). Both sources are detected
independently in both left and right-hand circular polarisations. 

\vspace{0.2cm} 
{\it
  Note that these two sources, imaged and detected via the same data
  set are separated by $\sim 2$~arcminutes on the sky - simultaneous
  detection of multiple sources in a single VLBI field of view will
  soon be the ``rule'' rather than the exception!  }. 
\vspace{0.2cm} 

There is a $4\sigma$ detection of J123646+621404, located within 20~mas
of the measured MERLIN position.  

Two other sources were not detected:
J123646+621448 and J123652+621444.

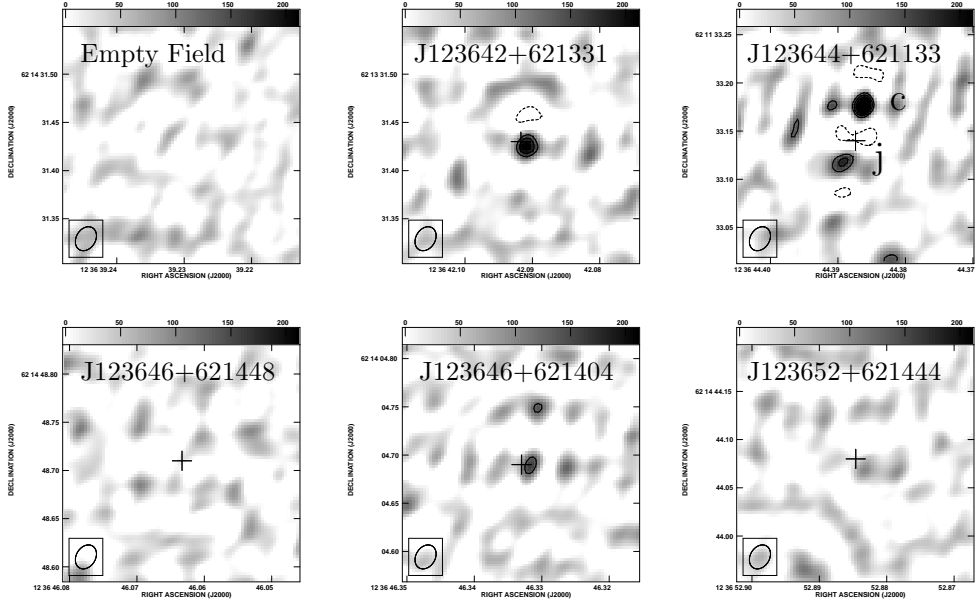
\begin{figure*}
\label{evn_hdf} 
\vspace{6cm}
\begin{picture}(150,100)

\put(0,0){\includegraphics{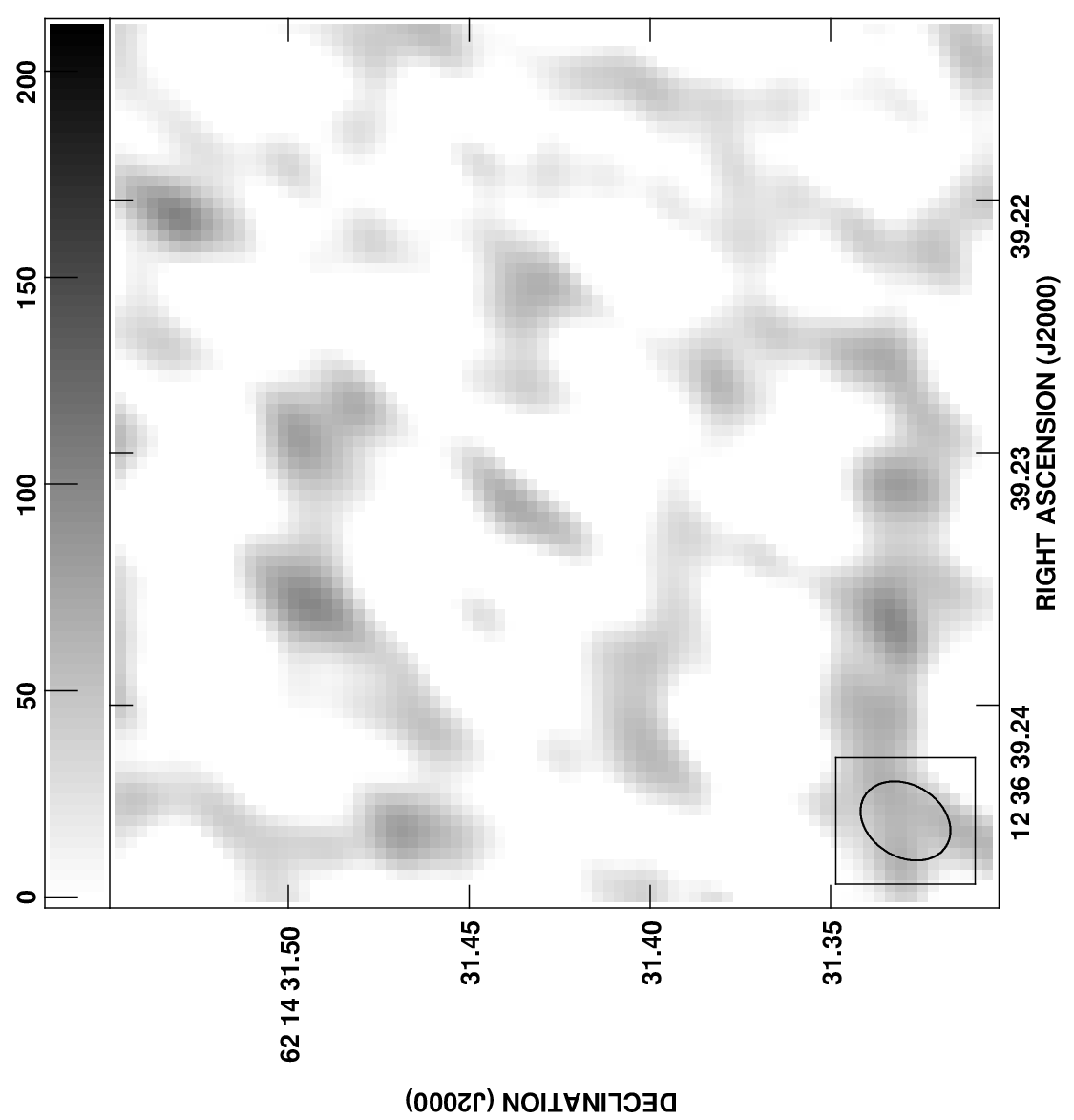}}
\put(36,214){Empty Field}

\put(0,0){\includegraphics{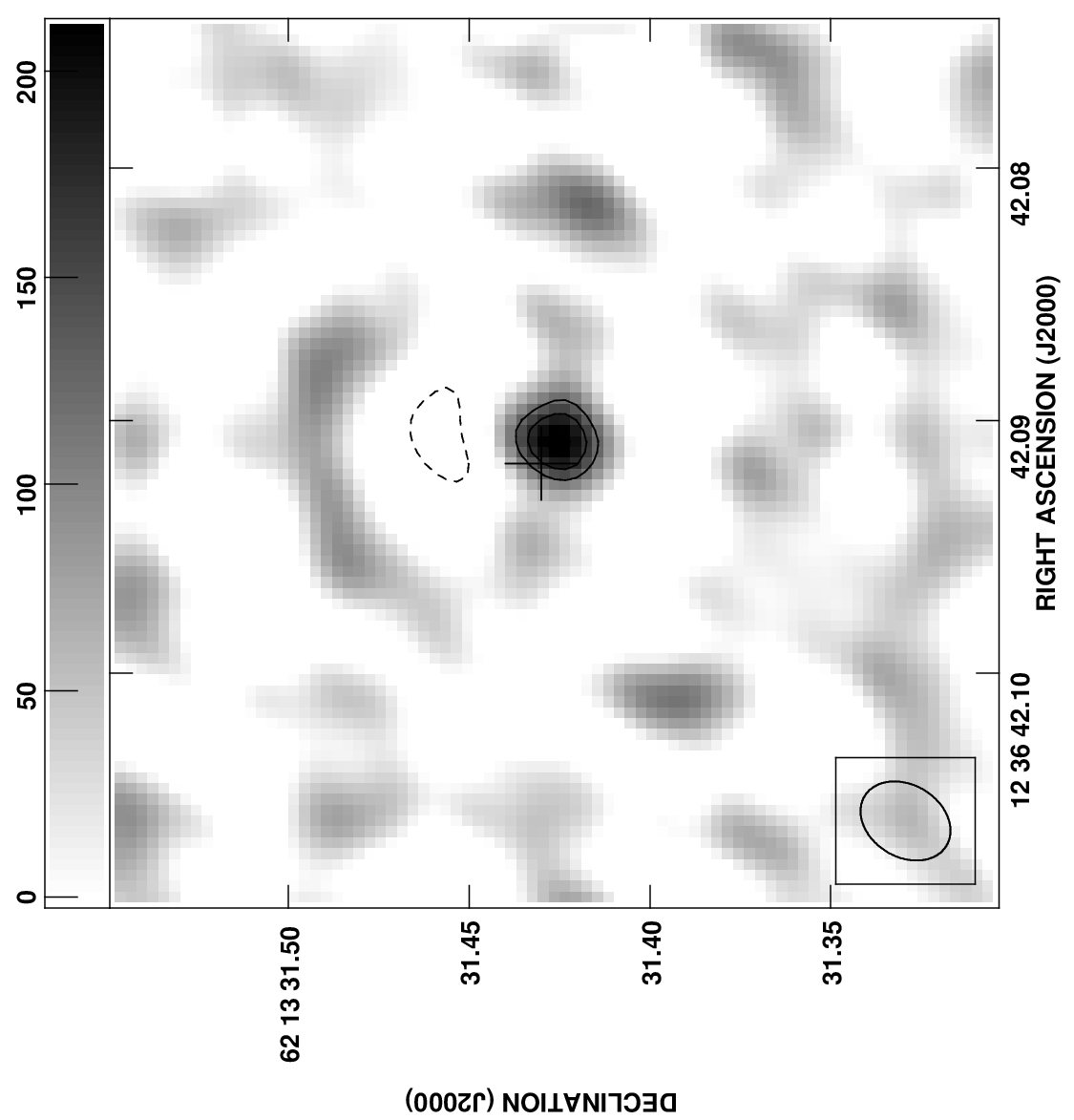}}
\put(162,214){J123642+621331}

\put(0,0){\includegraphics{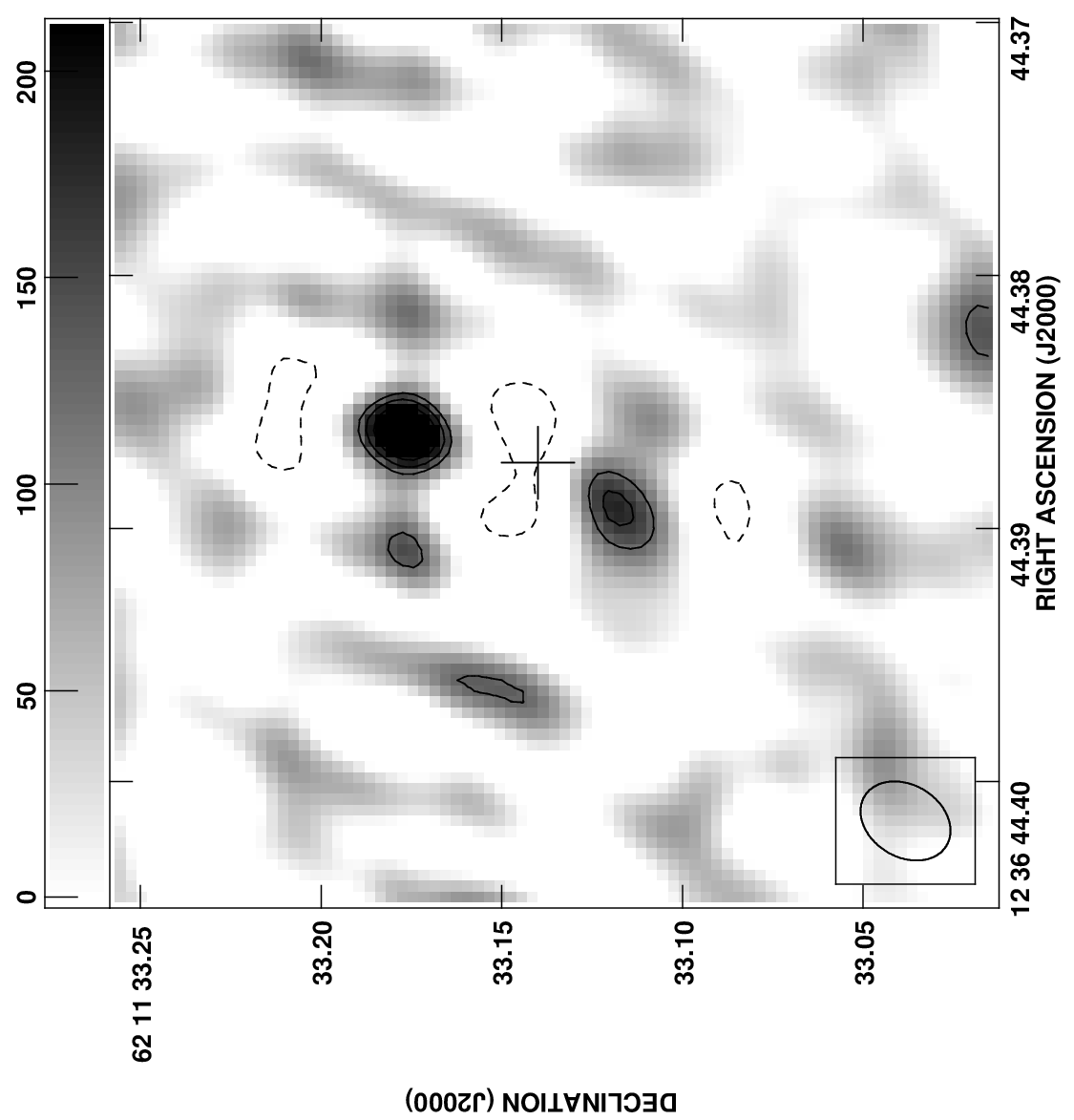}}
\put(288,214){J123644+621133}
\Large
\put(342,196){c}
\put(336,174){j}
\normalsize

\put(0,0){\includegraphics{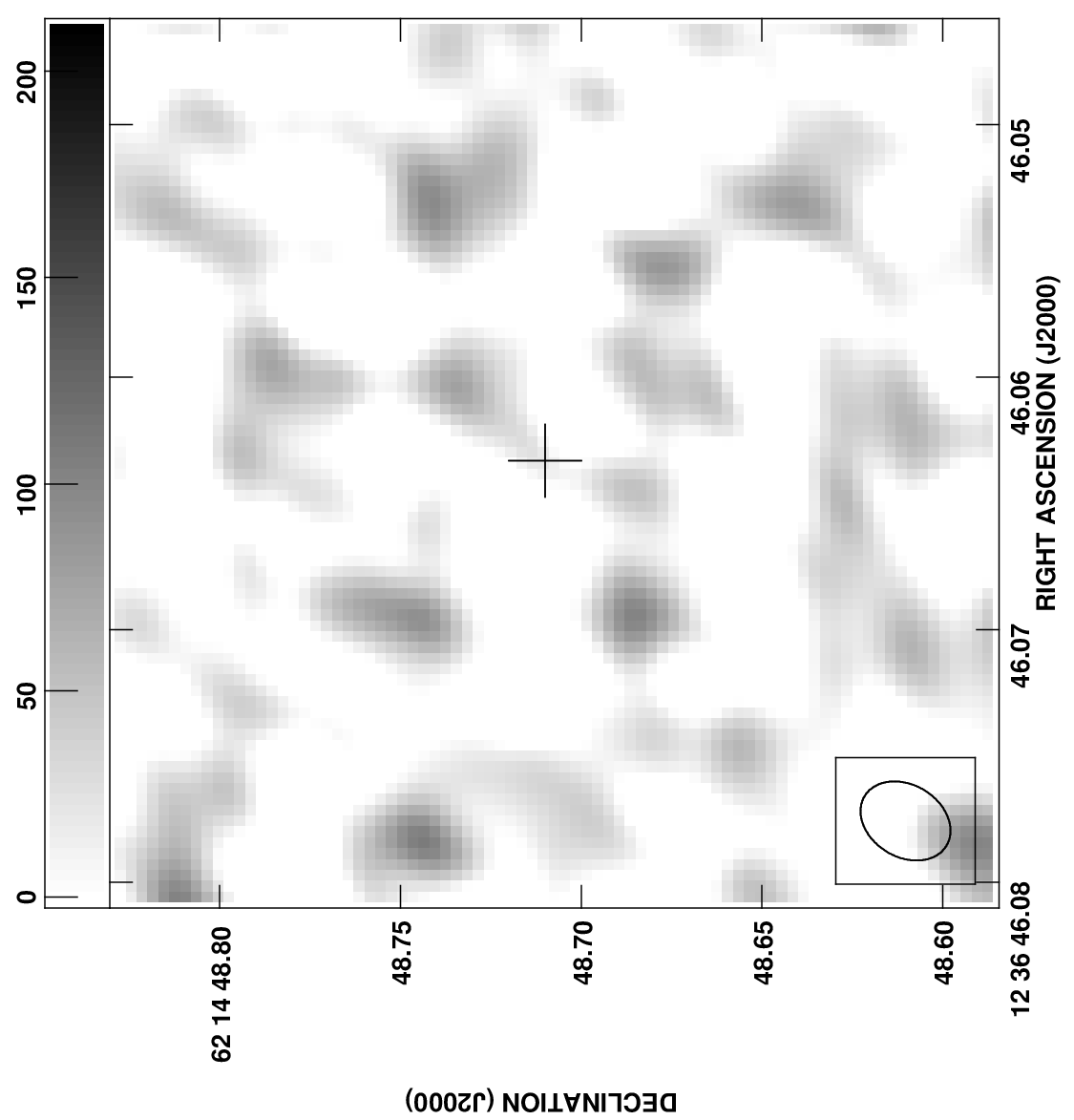}}
\put(36,94){J123646+621448}

\put(0,0){\includegraphics{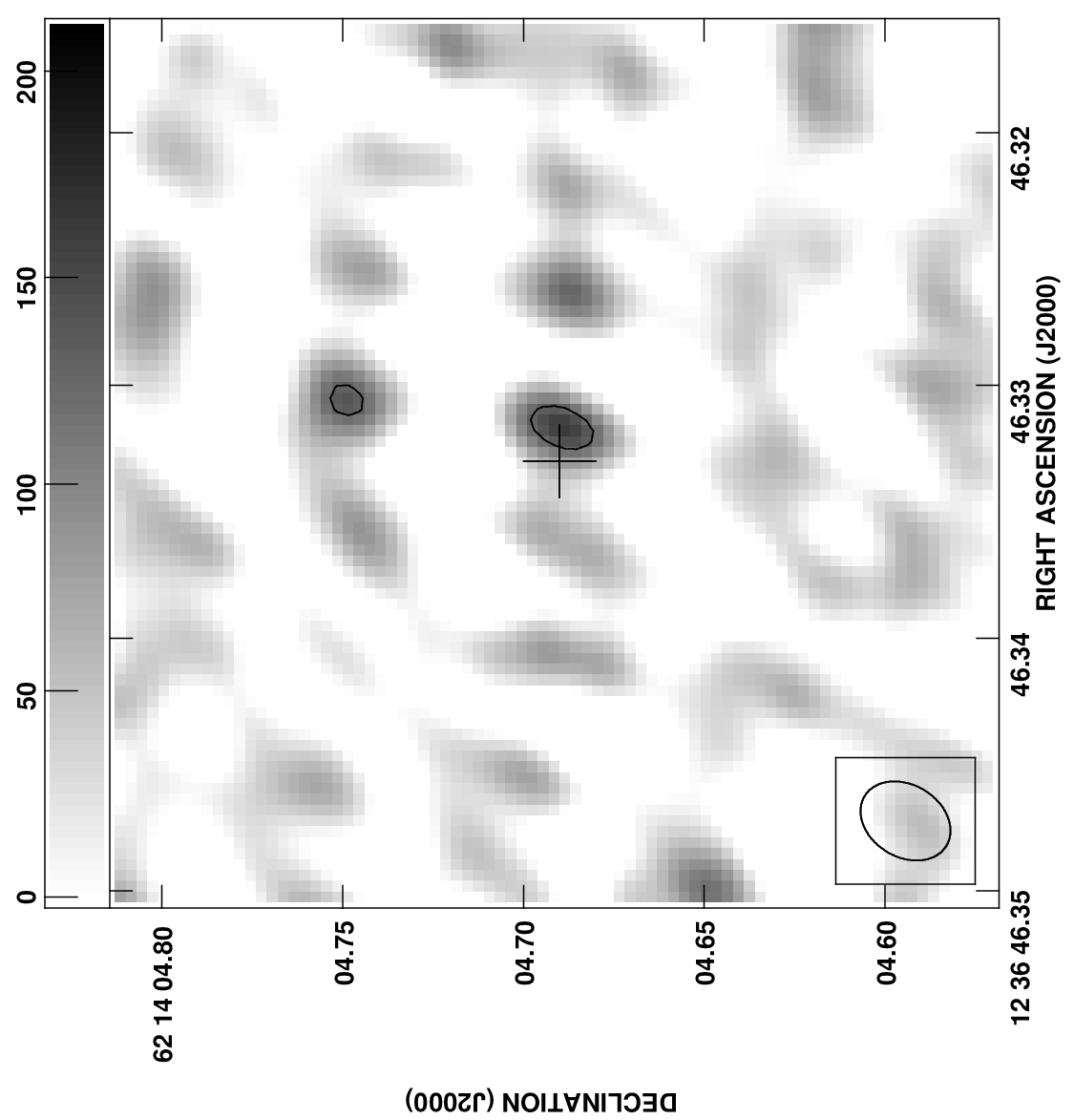}}
\put(164,94){J123646+621404}

\put(0,0){\includegraphics{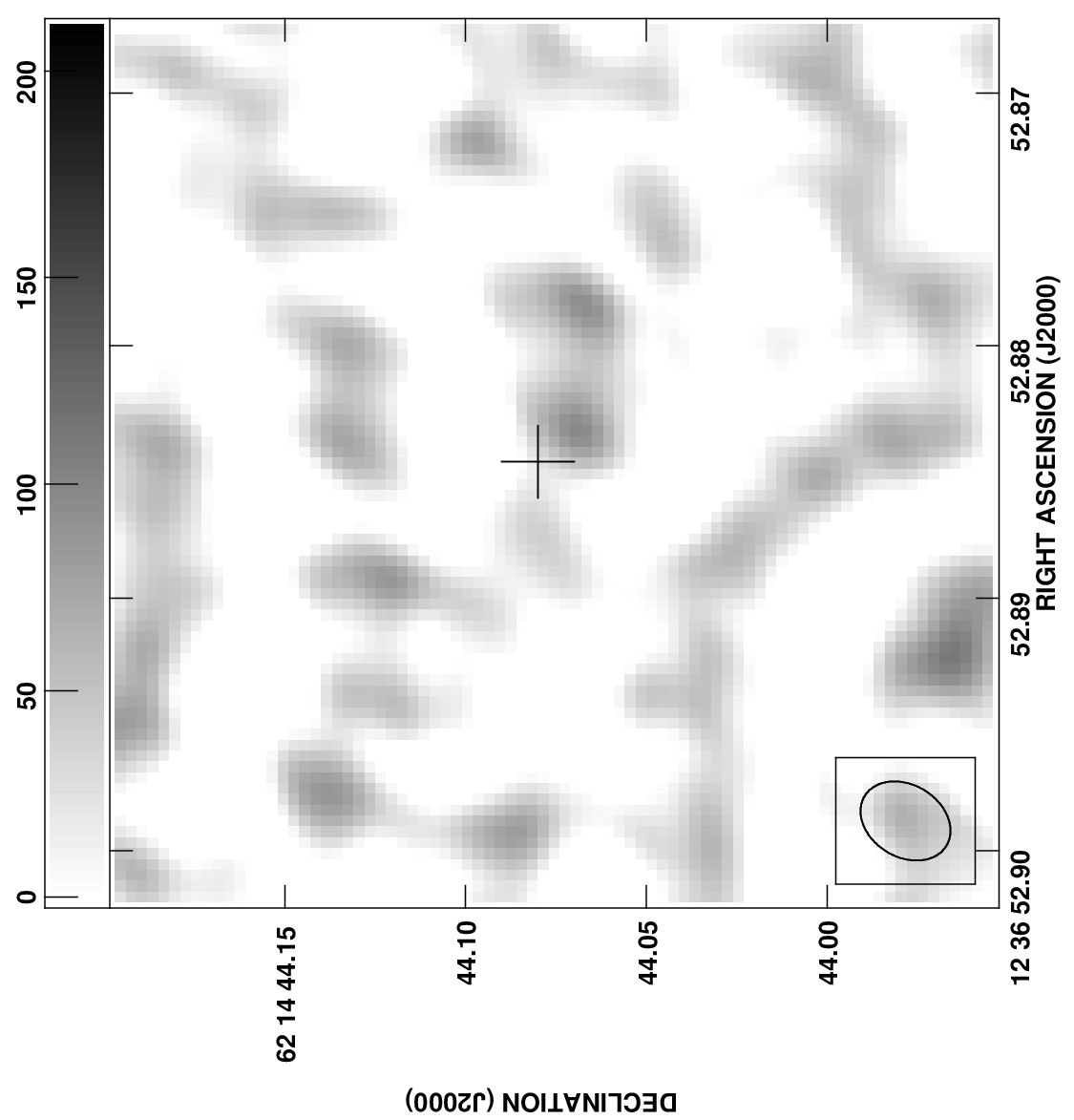}}
\put(288,94){J123652+621444}

\end{picture}
\caption{EVN 1.4 GHz dirty images of the six MERLIN targets. The
  MERLIN source positions are identified by crosses. Contours are 
drawn at -3, 3, 4, 5, 6, 7, 8, 9 times the rms noise level 
$\sim 40\mu$Jy per beam.}
\end{figure*}

\subsection{VLA~J123644+621133 and VLA~J123642+621331} 

VLA~J123644+621133 is identified with a bright ($I\sim 21.05^{m}$)
z=1.013, red elliptical galaxy (R98), with a radio luminosity and large
scale radio structure that classify it as an FR-I (R98).  The EVN (see
Fig.~1) detects a compact core component (c) with an estimated size $<
26$~mas, implying a brightness temperature, T$_{b} >
1.8\times10^{5}$~K. There also appears to be a clear $5\sigma$
detection of an additional component located $\sim 60$~mas south of the
core component (note that the MERLIN position lies roughly half-way
between components (c) and (j) -- the EVN thus provides {\it the} most
accurate (ICRF) source positions associated with the HDF. The
traditional VLBI structure observed in VLA~J123644+621133 supports the
view (see M99) that the rare ``classical'' radio sources detected in
the HDF and HFF, are simply related to the mJy tail of the AGN radio
source population.

VLA~J123642+621331 lies just outside the HDF, in an adjacent HFF. It is
a relatively bright, steep spectrum radio source
($S_{1.4GHz}\sim470\mu$Jy) with no optical counterpart to $I\leq
25^{m}$ (R98). Deep HST NICMOS $ 1.6\mu$m imaging clearly detects an
extremely red $23.9^{m}$ counterpart to the radio source (\cite{Wad99},
hereafter W99), and spectra obtained by the Keck~II telescope show a
single strong emission line, which is identified with Ly~$\alpha$ at
z=4.424. W99 interpret the source as a dust obscured, star-forming
galaxy with an embedded active nucleus.

The high resolution $0.15^{\prime\prime}$ 1.4~GHz VLA-MERLIN
observations (M99) show that approximately 10\% of the total radio
emission in VLA~J123642+621331 resides in an extended component, lying
to the east of an unresolved core. It is the core which is detected by
the EVN at 1.6~GHz. The main question is whether the radio emission
arises due to star-formation processes or AGN activity.  The EVN limit
on the measured core size ($< 26$ mas) implies a brightness temperature
$>10^{5}$K. The detection of this compact radio source, together with
its high luminosity of $\sim 10^{25}$W/Hz ($\sim 100$ times more
luminous than the luminous starburst Arp 220), argues strongly for an
AGN interpretation for most of the radio emission. A higher resolution
($21\times15$ mas), uniformly weighted image of VLA~J123642+621331
suggests that the core is resolved on the longer EVN baselines,
although it should be noted that the noise in this image is twice as
high as that in the naturally weighted case (see Fig.~1).  Further,
high resolution, global VLBI observations are clearly required to
confirm an AGN interpretation for this radio source.

\section{Conclusions}

\subsection{Results} 

We have clearly ($5\sigma$) detected two radio sources within the HDF
region.  One of these lies within the HDF itself (VLA~J123644+621133)
and shows all the usual properties of a low luminosity, classical FR-I
radio source.  The second, perhaps more interesting source is
VLA~J123642+621331. The EVN detection of a compact core supports the
suggestion that this is a dusty, star-forming galaxy in which the bulk
of the radio emission is associated with an embedded AGN.

\subsection{Future VLBI Deep Fields} 

Further high sensitivity Global VLBI observations of this source and
other sources in the HDF region are planned.  It may be noted that
significant short-term improvements in VLBI deep field imaging can be
expected.  A 2-fold increase in sensitivity is anticipated from the
MkIV system running at its full 1 Gbit/sec capacity. This, coupled with
multiple correlator passes associated with a judiciously chosen target
field (see \cite{G00b}), will permit VLBI r.m.s. noise levels to
routinely approach those currently enjoyed by connected element arrays.

\vspace{0.2cm}
{\it VLBI observers might want to bear in mind that embedded within
  every typical continuum VLBI data set, there are, in addition to the
  brighter target, 3 eminently detectable sub-mJy sources - all trying
  to get out!}
\vspace{0.2cm} 

Global VLBI observations with 3 times the resolution and 4 times the
sensitivity of the pilot observations presented here, are now
achievable. With a linear resolution of $\sim 30$ pc at cosmological
distances, only Global VLBI can resolve nuclear star-forming regions
like those observed in Arp~220 (\cite{S98}). Targeted observations of
the optically faint radio source population (of which
VLA~J123642+621331 is just one example) may throw new light on the
nature of the associated population of dusty sub-mm sources, recently
discovered by SCUBA (\cite{H98}).  In the absence of near-infrared, optical,
ultraviolet and even X-ray emission (due to dust obscuration), future
VLBI observations will be a crucial diagnostic in distinguishing
between AGN and starburst activity in these systems.

\end{document}